\documentclass{PoS}
\usepackage[utf8]{inputenc}
\usepackage[T1]{fontenc}
\usepackage{amsmath}
\usepackage{mathtools, subcaption}

\usepackage{wrapfig}
\usepackage{multicol}

\makeatother
\let\oldbibliography\thebibliography
\renewcommand{\thebibliography}[1]{\oldbibliography{#1}
	\setlength{\itemsep}{-.5pt}} 

\graphicspath{{./plots/}}

\newcommand{\bes}{\begin{eqnarray}}
\newcommand{\ees}{\end{eqnarray}}

\newcommand{\mqh}{m_{{\rm q},{\rm h}}}
\newcommand{\mql}{m_{{\rm q},{\rm l}}}
\def\mbar{\kern1pt\overline{\kern-1pt m\kern-1pt}\kern1pt}
\newcommand{\kaph}{\kappa_{\rm h}}

\def\gbar{\bar{g}}

\def\nf{N_{\rm f}}

\def\zm{Z_{\rm m}}
\def\za{Z_{\rm A}}
\def\zp{Z_{\rm P}}

\def\bm{b_{\rm m}}
\newcommand{\Rm}{R_{\rm m}}

\def\fm{{\rm fm}}
\def\MeV{{\rm MeV}}

\title{$\mathcal{O}(a)$ improved quark mass renormalization for a non-perturbative matching of HQET to three-flavor QCD}

\ShortTitle{Improved quark mass renormalization for a matching of HQET to three-flavor QCD}

\author{Patrick Fritzsch \hfill\parbox{28mm}{\vspace{-0.25cm}\raggedleft\footnotesize\it%
		CERN-TH-2018-236}\hspace{-9mm}\hspace{.1\textwidth}
	\\
	Theoretical Physics Department, CERN, \hfill\parbox{17.5mm}{\vspace{-0.25cm}\raggedleft\footnotesize\it%
		MS-TP-18-26}\\
	1211 Geneva 23, Switzerland\\
	E-mail: \email{patrick.fritzsch@cern.ch}}

\author{Jochen Heitger, \speaker{Simon Kuberski}\\
	Westfälische Wilhelms-Universität Münster, Institut für Theoretische Physik,\\
	Wilhelm-Klemm-Straße 9, 48149 Münster, Germany\\
	E-mail: \email{heitger@uni-muenster.de}, \email{simon.kuberski@uni-muenster.de} }

\abstract{The use of Heavy Quark Effective Theory (HQET) on the lattice as an approach to B-physics phenomenology is based on a non-perturbative matching of HQET to QCD in finite volume. As a first step to apply the underlying strategy in the three-flavor ($N_f = 2+1$) theory, we determine the renormalization constant and improvement coefficients relating the renormalized current and subtracted quark mass of (quenched) valence quarks in $\mathcal{O}(a)$ improved $N_f=3$ lattice QCD. We present our strategy and first results for the relevant parameter region towards weak couplings along a line of constant physics, which corresponds to lattice resolutions $a\leq 0.02\,$fm and fixes the physical extent of the matching volume to $L\approx 0.5\,$fm.}

\FullConference{The 36th Annual International Symposium on Lattice Field Theory - LATTICE2018\\
		22-28 July, 2018\\
		Michigan State University, East Lansing, Michigan, USA.}

\begin{document}
\section{Introduction: Matching of HQET to QCD}
We work along the lines of the ALPHA collaboration's strategy for B--physics via non-perturbative HQET, as it was introduced in \cite{Heitger:2003nj} and applied e.g. in \cite{Blossier:2010jk} and \cite{Blossier:2012qu}. Eventually, the results from HQET observables on the large volume CLS configurations \cite{Bruno:2014jqa} have to be matched with finite volume QCD calculations to determine the matrix elements we want to investigate.

The matching ensembles on the QCD side are generated on a so-called line of constant physics (LCP), to ensure clean extrapolations to the continuum at all stages of the computation. The first part of the definition of the LCP is a fixed physical extent $L$ of the generated lattice ensembles since the renormalization scale is defined by $\mu\equiv1/L$. The exact value of the lattice size is not of concern, as long as it is the same for all ensembles. For the matching volume, we aim for a lattice extension of about $L_1\approx0.5\,\fm$ which can be translated in a gradient flow coupling of $\gbar^2_\mathrm{GF}(L_1/2)\approx4$ \cite{DallaBrida:2016kgh}. As second part of the definition, we set the dynamical light quark masses to zero, $Lm(L)=0$ for our mass-independent renormalization scheme.

\vspace{1em}\noindent\textbf{Fixing the matching volume \& target precision}\\
\indent We wish to fix the physical extent of the matching volume $L_1$
through a careful tuning of the gradient flow running coupling 
$\gbar^2_\mathrm{GF}(L_1/2)$ at the smaller tuning volume with $L=L_0\equiv L_1/2$ and $T=L$. 
We aim at a relative precision of about $\Delta L_0/L_0 = 0.01$,
which amounts to an improvement of about a factor three compared to the $\nf=2$ work \cite{Blossier:2012qu}.

To translate this goal into an estimate for the target precision of the coupling,
one can employ the renormalization group equation, which implies for
generic $L$ and associated finite-volume renormalized coupling $\gbar(L)$ 
and $\beta$-function $\beta(\gbar)$, respectively,
\bes
\frac{\Delta L}{L} =
-\frac{\Delta\gbar^2}{2\,\gbar\,\beta(\gbar)} =
\left[\frac{-\gbar}{2\,\beta(\gbar)}\right]\frac{\Delta\gbar^2}{\gbar^2}
\label{e:relprec_L}
\ees
in terms of the relative error $\Delta\gbar^2/\gbar^2$ of $\gbar^2$.

Since the envisaged matching volume corresponds to values of the renormalized coupling
$\gbar\equiv\gbar_\mathrm{GF}$ in the non-perturbative regime, we employ for
$\beta(\gbar)$ the expression taken from
eq.~(4.12) of \cite{DallaBrida:2016kgh}, $\beta(\gbar) =
-{\gbar^3}/\left(p_0+p_1\gbar^2+p_2\gbar^4\right),$
and the choice (4.15) for the coefficients from the same reference, 
which parametrizes the non-perturbative $\beta$-function as determined from
the results on the gradient flow running coupling in this low-energy regime. 

The target for the gradient flow coupling $\gbar^2_\mathrm{GF}(L_1/2)=\gbar_*^2$
is set by our largest ensemble with $L_0/a=32$ lattice points in all spatial and temporal
directions. After the coupling on this ensemble has been measured to the desired precision, 
the ensembles with larger lattice spacings are tuned to this target coupling. 
As a guide for the initially guessed	 $\beta\equiv6/g_0^2$ for the ${L_0/a=24,32}$ ensembles,
the available coupling data in table~1 of \cite{DallaBrida:2016kgh} was used. It served also 
as starting point for interpolations in $\beta$ to reach the desired coupling. The current status
of the tuning is shown in fig.~\ref{f:tuning_couplings}.
 
With $\gbar_*^2=\gbar^2_\mathrm{GF}(L_0/a=32)=3.949$ as the value determined on our simulation
with ${L_0/a=32}$, one can solve
(\ref{e:relprec_L}) for $\Delta\gbar^2/\gbar^2$. One then arrives at 
$\frac{\Delta\gbar^2}{\gbar^2} = 0.0047 \,$ and $\Delta\gbar^2 = 0.0187 \,$
for the relative and absolute errors as the desired target 
precision of the gradient flow coupling at the target coupling.
The use of the perturbative $\beta$-function with the two universal coefficients $b_0$ and $b_1$ from ref.~\cite{Campos:2018ahf} leads to slightly larger values.

Having the final value of the coupling at hand, 
we can also estimate the extent of the volume in physical units that is
implicitly fixed by a value of the coupling using the results of \cite{Bruno:2017gxd}.
This is done applying eq.~(5.5) of \cite{DallaBrida:2016kgh} for the scale 
factor $s=L_2/L_1$ between two given coupling values, $g_2=\gbar(L_2)$ and 
$g_1=\gbar(L_1)$, and inserting the non-perturbative parametrization of $\beta$ from
above.

The physical extent then can be calculated with the additional knowledge of the value of the coupling at a certain
hadronic scale corresponding to the choice "$\mu_{{\rm had},1}$" in \cite{Bruno:2017gxd}.
This is related through ${\mu_{\rm ref}^{\star}}/{\mu_{{\rm had}}} = 2.428(18)$
to the reference scale $\mu_{\rm ref}^{\star} = 478(7)\,\MeV,$ which 
is defined and extracted from the CLS $\nf=2+1$ large volume simulation results in \cite{Bruno:2017gxd}.
With $L_0=\frac{2.428}{s}\,L_{\rm ref}\,,$
the scaling factor $s$ and the couplings ${\gbar^2_\mathrm{GF}(L_{\rm had}) =11.31 }\,$ and ${\gbar^2_\mathrm{GF}(L_0)=3.949}\,,$
we get
\bes
L_0 = 0.25\,\fm \,,\qquad
L_1=2L_0 = 0.50\,\fm \,.
\ees
for the physical extents of our ensembles. In the matching volume $L_1$ we use five different resolutions with lattices ranging from 24 to 64 lattice points in every direction corresponding to lattice spacings $a$ with $0.0078\,\mathrm{fm} \leq a \leq 0.021\,\mathrm{fm}$. 

\begin{figure}
	\begin{minipage}{.45\textwidth}
		\hspace{-.05\textwidth}
		\input{./plots/couplings.tex}
		\caption{Current status of the tuning: Gradient flow coupling $\gbar^2_\mathrm{GF}(L_1/2)$ for the tuned ensembles with varying lattice spacings.
			\label{f:tuning_couplings}}
	\end{minipage}
	\hspace{.05\textwidth}
	\begin{minipage}{.42\textwidth}
		\input{./plots/masses_fit_12.tex}
		\caption{Fit of $m_\mathrm{hh}$ and $m_\mathrm{lh}$ to the PCAC masses versus ${\Delta am_\mathrm{h}}$ for ${L_0/a=12}$. Errors are smaller than the symbol sizes.
			\label{f:masses_fit}}
	\end{minipage}
\end{figure}

\section{Improvement coefficients and renormalization constant: Strategy}
The computation of the dependence of the finite-volume observables on the renormalized heavy quark mass requires the determination of various improvement coefficients and renormalization constants. The renormalization group invariant (RGI) mass in the $\mathcal{O}(a)$ improved theory is determined from the bare subtracted heavy quark mass $\mqh$ via the relation 
\begin{align}
M_{\rm h}\;=\;
h(L)\,\zm(g_0,L/a)\,\left(1+\bm(g_0)\,a\mqh\right)\,\mqh
\,+\,\mathcal{O}(a^2)\,, \label{e:RGI_mass}\\[1em]
\mathrm{where}\qquad\zm(g_0,L/a)=
\frac{Z(g_0)\,\za(g_0)}{\zp(g_0,L/a)}\;,\qquad
a\mqh=
\frac{1}{2}\left(\frac{1}{\kaph}-\frac{1}{\kappa_{\mathrm{cr}}}\right)\;.
\end{align}
The renormalization factor of the axial vector current $\za$ is available from \cite{Bulava:2016ktf} and \cite{DallaBrida:2018tpn}, and the necessary interpolation formula for the running factor $h(L)$ is given in \cite{Campos:2018ahf}.
With these formulas, we end up with $h(L_0)={M}/{\mbar(L_0)} = 1.4744(87)\,$
for the running factor to our tuning volume.

For the calculation of the RGI mass following eq.~(\ref{e:RGI_mass}), the improvement coefficient $\bm$ and the normalization constants $Z$ and $\zp$ need to be determined. These quantities are calculated on the tuning ensembles with $L=L_0$ to avoid correlations with the matching observables.

The pseudoscalar renormalization constant $\zp$ can be calculated following the strategy of \cite{hep-lat/0507035}. The definitions of estimators to calculate the improvement coefficient $\bm$  and the renormalization constant $Z$ based on the non-degenerate improved bare current quark mass $m_{ij}$  from the PCAC relation have first been derived in the context of the quenched analysis in \cite{Guagnelli:2000jw}.

As it can be seen, in eqs.~(2.26--2.28) of \cite{Fritzsch:2010aw}, $\Rm$ can be calculated up to $\mathcal{O}(a)$ effects which introduces $\mathcal{O}(a^2)$ effects in the masses, whereas $R_Z$, to this order, suffers only from sea quark effects. While at the chosen line of constant physics the sea quark effects are expected to vanish, the $\mathcal{O}(a)$ ambiguities due to heavy valence quark masses can be substantial.

In contrast to the quenched and the two-flavor case, the estimators are not directly calculated for distinct choices of degenerate and non-degenerate valence quark masses. Instead they are determined from smoothly interpolated functions $m_\mathrm{hh}(x)$ of the degenerate heavy-heavy masses and $m_\mathrm{lh}(x)$ of the non-degenerate light-heavy masses, depending on the mass difference
\begin{align}
	x \equiv \Delta am_\mathrm{h} = a\mqh - a\mql = \frac{1}{2}\left(\frac{1}{\kappa_\mathrm{h}}-\frac{1}{\kappa_\mathrm{l}}\right)
\end{align}
in the subtracted quark masses, where l denotes the light valence quark mass equal to the sea quark masses.
The used fit formulas for the heavy-heavy $m_\mathrm{hh}(x)$ and the light-heavy current masses $m_\mathrm{lh}(x)$
and the resulting formulas for the estimators are justified and derived in full detail in \cite{DeDivitiis2018}, where our fit method is established for a region of stronger coupling compared to the case described here.

In ref.~\cite{Fritzsch:2010aw} it was observed, that the $\mathcal{O}(a)$ ambiguities are suppressed, when the improvement coefficients are calculated at the hopping parameter of the bare subtracted quark mass, which is improved. The interpolation strategy makes it possible to determine the coefficients at this hopping parameter at the stage of the matching, after the measurements have been finished.

\section{Improvement coefficients and renormalization constant: Results} 
\noindent\textbf{Interpolation in the heavy quark mass}\\
\indent We applied our strategy for the extraction of the improvement coefficients and the renormalization constant to five different ensembles with varying lattice spacing, all lying on the LCP defined above. We use the Schrödinger functional setup on $L^3\times T$, $L=T$, lattices with Lüscher-Weisz gluons and three degenerate flavors of Wilson-clover fermions. On every ensemble the necessary Schrödinger functional correlation functions for the calculation of the PCAC masses are measured for several heavy valence quarks on a range between the (vanishing) sea quark mass and the bottom quark mass. For some ensembles measurements at negative quark masses are done, in order to investigate their effect on stabilizing the fits.
\begin{figure}
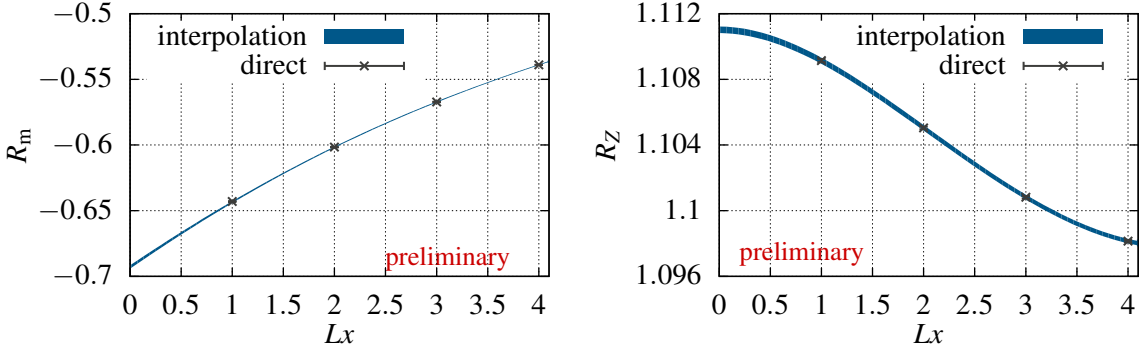

	\begin{subfigure}[c]{0.475\textwidth}
		\input{./plots/rM_cuw.tex}
		\label{f:rM_cuw}
	\end{subfigure}
	\hspace{0.025\textwidth}
	\begin{subfigure}[c]{0.475\textwidth}
		\input{./plots/rZ_cuw.tex}
		\label{f:rZ_cuw}
	\end{subfigure}
	\vspace{-18pt}
	\caption{Estimators $\Rm$ and $R_Z$ based on the interpolation in the current quark masses ${Lx = \frac{L}{2}(\frac{1}{\kappa_\mathrm{h}}-\frac{1}{\kappa_\mathrm{l}})}$ together with the estimators based on the plateau averages of calculations on single time slices.\label{f:rxZ_cuw}}
\end{figure}

A representative result for the simultaneous fit of the functions $m_\mathrm{hh}(x)$ and $m_\mathrm{lh}(x)$ to the calculated PCAC masses is shown in figure~\ref{f:masses_fit}. As constructed by the fit functions, the curves meet at the unitary point $\rm l= \rm h$ at $x=0$. Having the interpolation formulas at hand, the estimators can be calculated for every heavy valence quark mass inside our interpolation range. 

From the definitions of the estimators, variations of $\mathcal{O}(a\mqh)$ are expected. To investigate these variations, the estimators are calculated in the whole chosen range of valence quark masses. The results for the coarsest ensemble are shown in figure~\ref{f:rxZ_cuw}, where the errors are the statistical ones calculated with the $\Gamma$-method \cite{Wolff:2003sm}, to handle possible autocorrelations. As expected, the estimator $R_Z$ shows only mild variations with the heavy quark mass, while for $\Rm$ the variations are quite significant. The estimators coming from the interpolating functions can be compared with estimators directly calculated in the way it was done in the former analyses \cite{Guagnelli:2000jw,Fritzsch:2010aw}. As expected, the directly calculated values fall onto the interpolating curves, which gives us confidence in the correctness of our method.

\begin{figure}
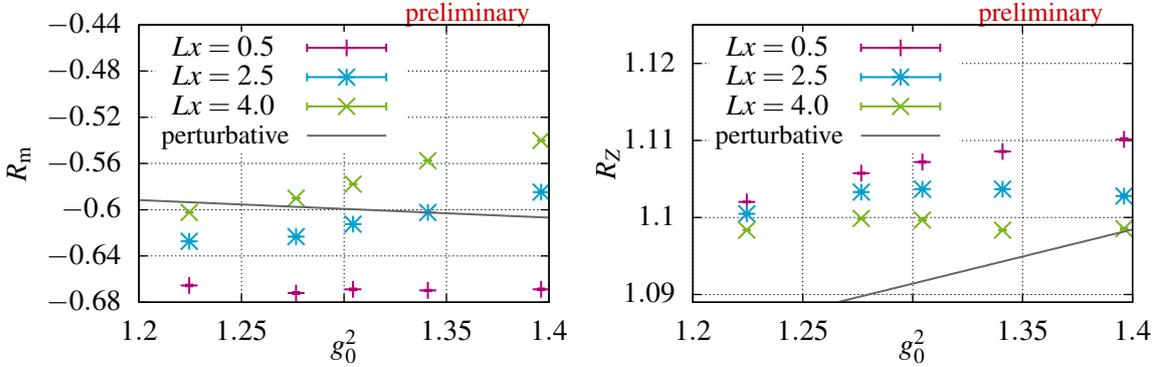

	\begin{subfigure}[c]{0.475\textwidth}
		\input{./plots/rm_coupling_proc.tex}
		\label{f:rM_coupling}
	\end{subfigure}
	\hspace{.02\textwidth}
	\begin{subfigure}[c]{0.475\textwidth}
		\input{./plots/rz_coupling_proc.tex}
		\label{f:rZ_coupling}
	\end{subfigure}
	\vspace{-18pt}
	\caption{Coupling dependence of the estimators $\Rm$ and $R_Z$ for three different choices of heavy quark masses together with the one-loop perturbative predictions from \cite{Taniguchi:1998pf} and \cite{Aoki:1998ar}. \label{f:RX_coupling}}
\end{figure}
\vspace{1em}\noindent\textbf{Dependence on the coupling}\\
\indent When the analysis is done for all lattice spacings, the dependence of the estimators on the bare coupling $g_0^2$ can be investigated. The LCP can now be extended to include the condition $Lx=\mathrm{const.}$ for the valence quarks to ensure the same physics on every ensemble. This fixation to the evaluation at a constant mass difference does not need a fine-tuning of the valence quark hopping parameters $\kappa_{\mathrm{h}}$, as it would have been the case for a determination at distinct mass differences. The results for three different choices of heavy valence quark masses are depicted in figure~\ref{f:RX_coupling} together with the one-loop perturbative predictions. The estimators seem to vary smoothly with the coupling, slight deviations may have their origin in the fact that at the shown status the used ensembles were not fully tuned to the line of constant physics. As expected, the differences between the three chosen quark masses get smaller, when the bare coupling (and with it the lattice spacing) is decreased. 

For the improvement and renormalization of the quark mass on our matching ensembles, the results from the tuning ensembles can be used directly. For future use, interpolation formulas for the shown coupling region will be fitted to the final data, as it was done in the two-flavor case \cite{Fritzsch:2010aw} and the CLS coupling region \cite{DeDivitiis2018}. In any case the position of the non-perturbatively determined points and the curvature of the $g_0^2$-behavior show a considerable deviation from the perturbative prediction and it remains open, how the curves approach in the limit of vanishing coupling.

\vspace{1em}\noindent\textbf{Investigation of $\mathcal{O}(a)$ ambiguities}\\[-\the\baselineskip] 
\begin{wrapfigure}{r}{0.47\textwidth}
	\vspace{-.9cm}
	\hspace{-.5cm}
	\input{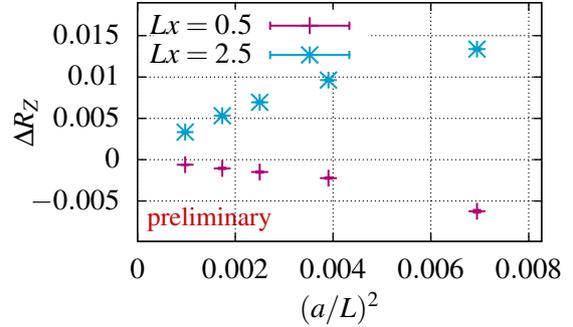}
	\caption{Differences $\Delta R_Z$ between standard and improved derivatives towards the continuum limit.\label{f:RX_deviation}}
	\vspace{-1em}
\end{wrapfigure} 
\indent The estimators only give the improvement coefficient up to $\mathcal{O}(a)$ and (for vanishing sea quark mass) the renormalization constant up to $\mathcal{O}(a^2)$. Different choices in the improvement conditions lead to different $\mathcal{O}(a)$ effects, i.e. there is an ambiguity between the possible choices. Every change in the improvement conditions gives a new set of estimators. This does not introduce any problems in the continuum extrapolation, as long as the same condition is chosen for every lattice spacing, since the differences are expected to vanish smoothly in the continuum \cite{Guagnelli:2000jw}. 

The expected disappearance of the ambiguities in the continuum limit is investigated by comparing different choices in the improvement conditions. Here, the definitions of the lattice derivatives for the calculation of the PCAC masses is changed as it was done in ref.~\cite{Guagnelli:2000jw}.

The differences ${\Delta R_Z = R_Z(\text{impr. deriv.}) - R_Z(\text{std. deriv.})}$
are depicted in figure~\ref{f:RX_deviation}. Towards the continuum limit, the anticipated $\mathcal{O}(a^2)$ effects for $R_Z$ can be seen. Similar test have been done for various choices in the definition of our improvement condition and all estimators.

\section{Discussion and outlook}
With the values for $\bm$ and $Z$ calculated in this work, $\zp$ from the same ensembles, $h(L)$ from \cite{Campos:2018ahf} and $\za$ from \cite{DallaBrida:2018tpn}, the RGI mass $M_{\rm h}$ (and $z\equiv LM_{\rm h}$) can be calculated from eq.~(\ref{e:RGI_mass}). With the current status of the tuning runs, an error budget for the final results can be estimated from the finest lattice with $L_0/a=32$. The error on the running factor $h(L)$ is to be added in quadrature to the error from the other factors only in the continuum limit, cf. \cite{Campos:2018ahf}. Without the running factor and with the current statistics, we get a relative error $\Delta z / z = 0.31\,\%$. When the error on the running factor $\Delta h / h = 0.59\,\%$ is added, we end up with a total error of $\Delta z / z = 0.67\,\%$. As in the two-flavor case \cite{Fritzsch:2010aw}, the running factor is responsible for the dominant part of the total error, although the relative error on $h$ is reduced by 35\% compared to the two-flavor analysis \cite{hep-lat/0507035}. 

On the way to the non-perturbative finite-volume matching of QCD to HQET we made the first step by tuning the bare parameters of our ensembles in the tuning volume with $L=L_1/2$ to ensure a line of constant physics. The use of the gradient flow coupling enables us to tune the physical lattice sizes to the desired precision. Already existing interpolation formulas ensured a successful tuning to vanishing sea quark masses. By now, first simulations for our ensembles in the matching volume have been started with the parameters gained from the tuning on the small lattices. 

Although the tuning is nearly completed, the tuning ensembles are still of use. The missing factors for the calculation of the RGI heavy quark mass in the matching procedure can be calculated on these lattices. With the interpolations in the valence quark masses, we are able to determine the improvement coefficient and the renormalization constant at any heavy quark mass in a range from massless quarks to the bottom quark. 

Apart from the generation of the computational very costly QCD ensembles, also the HQET ensembles in the matching volume have to be generated. This is our next step towards the non-perturbative matching.\\

\textbf{Acknowledgments:} We thank R.\ Sommer for his work in our project on HQET, G.\ M.\ de Divitiis, C.\ C.\ Köster and A.\ Vladikas for their collaboration in our project in the CLS coupling region and F.\ Joswig for helpful discussions. This work is supported by the Deutsche Forschungsgemeinschaft (DFG) through the Research Training Group \textit{“GRK 2149: Strong and Weak Interactions – from Hadrons to Dark Matter”} (J. H. and S. K.). We gratefully acknowledge the Gauss Centre for Supercomputing e.V. (\href{www.gauss-centre.eu}{www.gauss-centre.eu}) for funding this project by providing computing time on the GCS Supercomputer SuperMUC at Leibniz Supercomputing Centre (\href{www.lrz.de}{www.lrz.de}). We acknowledge the computer resources provided by the \textit{ZIV} of the University of Münster (PALMA \& PALMA II HPC clusters) and thank its staff for support.

\section*{References}
\begin{multicols}{2}
	\small
	\renewcommand{\refname}{ \vspace{-\baselineskip}\vspace{-.68em} }
	
\end{multicols}

\end{document}